\documentstyle[12pt]{article}

\makeatletter

\@addtoreset{equation}{section}
\def\section{\@startsection {section}{1}{\z@}{-3.5ex plus -1ex minus
 -.2ex}{2.3ex plus .2ex}{\large\bf\centering}}
\def\subsection{\@startsection{subsection}{2}{\z@}{-3.25ex plus%
 -1ex minus -.2ex}{1.5ex plus .2ex}{\sc}}

\advance \voffset by -1.0in
\advance \hoffset by -0.6in
\def\tr{\mbox{tr}}
\def\cd{\!\cdot\!}

\def\bea{\begin{eqnarray}}
\def\eea{\end{eqnarray}}

\def\bphi{{\mbox{\boldmath $\phi$}}}
\def\bvphi{{\mbox{\boldmath $\varphi$}}}

\def\bsigma{{\mbox{\boldmath $\sigma$}}}

\def\bep{{\mbox{\boldmath $\epsilon$}}}

\def\bp{{\mbox{\boldmath $p$}}}
\def\bj{{\mbox{\boldmath $j$}}}
\def\bd{{\mbox{\boldmath $d$}}}
\def\be{{\mbox{\boldmath $e$}}}

\def\bs{{\mbox{\boldmath $s$}}}

\def\bq{{\mbox{\boldmath $q$}}}
\def\bn{{\mbox{\boldmath $n$}}}

\def\bx{{\mbox{\boldmath $x$}}}

\def\bR{{\mbox{\boldmath $R$}}}

\def\bD{{\mbox{\boldmath $D$}}}

\def\b0{{\mbox{\boldmath $0$}}}

\begin{document}
\baselineskip 18pt
\parskip 7pt
\begin{flushright}

DTP 94-23

22 June 1994

\end{flushright}

\vspace {1.5cm}

\begin{center}

{\LARGE Multisolitons in a Two-dimensional Skyrme Model}

\vspace{0.6cm}
\baselineskip 24pt

{\Large
B.M.A.G. Piette$^1$, B.J.Schroers$^2$ and
 W.J.Zakrzewski$^{3,4}$
\\
Department of Mathematical Sciences, South Road
\\
Durham DH1 3LE, United Kingdom}
\\
\baselineskip 18pt
$^1${e-mail: {\tt b.m.a.g.piette@durham.ac.uk}}\\
$^2${e-mail: {\tt b.j.schroers@durham.ac.uk}} \\
$^3${e-mail: {\tt w.j.zakrzewski@durham.ac.uk}}\\
$^4$ also at the  Centre de Recherche Math\'ematiques, Universit\'e de
Montreal,
Canada

\vspace{1.5cm}

{\bf Abstract}

\end{center}
\baselineskip 15pt

{\small
The Skyrme model can be generalised to a situation where static
fields are maps from one Riemannian manifold to another. Here we study
a Skyrme model where
physical space is   two-dimensional euclidean space  and the target space  is
the two-sphere with its standard metric. The model has topological soliton
solutions  which are exponentially localised.
We describe a superposition procedure for solitons in our model  and derive an
 expression for the interaction potential of two solitons which only
involves the solitons' asymptotic  fields. If  the solitons have  topological
degree 1 or 2 there are  simple formulae for their interaction  potentials
which we use to
prove the existence of solitons of higher degree. We explicitly compute the
fields and
energy distributions for solitons  of degrees between  one
and six  and
discuss their geometrical shapes and binding energies. }

\baselineskip 18pt

\section{A Skyrme Model in Two Dimensions}

The Skyrme model is a non-linear  theory for $SU(2)$ valued fields in 3
(spatial) dimensions which
has soliton solutions. Each soliton has an associated integer topological
charge or degree which Skyrme identified with the baryon number \cite{Skyrme}.
 A soliton
with topological charge one is called a Skyrmion; suitably quantised it
is a model for a physical nucleon. Solitons of higher topological charge,
called multisolitons, are classical models for higher nuclei.

In this paper we study multisolitons in a two-dimensional version of the
Skyrme model. The model was first considered in \cite{Tigran}, but the
motivation there  is  somewhat different from the  approach taken here.
For the present  purpose it is important to be clear  in what sense our model
resembles
 Skyrme's  model. In this section we will therefore briefly review a
general framework for the  Skyrme model due to Manton \cite{M}
and explain  how our model fits into that  framework.
In \cite{M} the usual Skyrme energy functional is interpreted in terms of
elasticity
theory and
a  static Skyrme field is a map
\bea
\pi :S\mapsto \Sigma
\eea
from physical space $S$ to the target space $\Sigma$. Both $S$ and $\Sigma$
are assumed to be Riemannian manifolds with metrics $t$ and $\tau$
respectively.
The energy of a configuration $\pi$ is expressed in terms of its strain tensor
$D$. To calculate the strain tensor one introduces coordinates
$p^i$ on $S$ and $\pi^{\alpha}$ on $\Sigma$ and orthonormal frame fields
$\bs_m$ on $S$ and $\bsigma_{\mu}$ on $\Sigma$ ($1\leq i,m\leq \mbox{dim} S,\,
1\leq \alpha,\mu \leq \mbox{dim}\Sigma$). The  Jacobian of the
map $\pi$ is, in orthonormal coordinates,
\bea
J_{m\mu}= s_m^i{\partial \pi^{\alpha}\over \partial p^i}\sigma_{\mu \alpha}
\eea
and the strain tensor $D$ is defined via
\bea
D_{mn}= J_{m\mu}J_{n\mu}.
\eea
The energy functional should be a function of the strain tensor but it should
not depend on  the choice of  the orthonormal frame $\bs_m$. The
basic invariants  under orthogonal transformations of $D$, however, are well
known:  they are  the coefficients in the characteristic polynomial
$\chi_D (t) :=$ det$(D-t$\,id$)$.
 If $S$ is three-dimensional there
are three such invariants,  namely $\tr D,\, {1\over 2}(\tr D)^2-{1\over 2}\tr
D^2$ and $\mbox{det} D$.
  It is explained in \cite{M} that in the
usual Skyrme model,   where $S ={\bf R}^3$ and $\Sigma =SU(2)$, the energy
functional is constructed from  the
first two:
\bea
E_{\mbox{\tiny Skyrme}}= \int d^3x \, \left(
\tr D + {1\over 2}(\tr D)^2-{1\over 2}\tr D^2 \right).
\eea

It  is now straightforward  to construct a Skyrme model in two dimensions  in
this
framework. We will reserve the term Skyrme model for Skyrme's original model
and refer to our two-dimensional model as a baby Skyrme model, its soliton
solutions baby Skyrmions etc.
We want to  work in flat space, so we choose $S={\bf R}^2$. The choice of
$\Sigma$ is less clear. In order to obtain solitons with a topological
charge we require, as we shall explain,  $\pi_2(\Sigma)= {\bf Z}$, so the
simplest choice is
$\Sigma = S^2$. Thus a baby Skyrme field is a map
\bea
\bphi: {\bf R}^2 \mapsto S^2,
\eea
where $S^2$ is the unit 2-sphere in euclidean 3-space with the metric induced
by
that embedding, and we think of $\bphi$ as a three component vector
$(\phi_1,\phi_2,\phi_3)$
satisfying $\bphi^2 =\phi_1^2+\phi_2^2+\phi_3^2 =1$.
In two dimensions there are only two invariants of the strain tensor, namely
\bea
e_2 & =& \tr D = \partial_1\bphi^2 +\partial_2\bphi^2 \nonumber \\
e_4 &=& \mbox{det} D =(\partial_1\bphi\times\partial_2\bphi)^2,
\eea
where $\partial_i, i=1,2,$ denotes the partial derivative with respect to the
cartesian coordinates $x^i$ of the  vector $\bx \in {\bf R}^2$ and $\times$  is
 the vector product in three dimensions.

The simplest choice of the  energy functional is $E_{\sigma} ={1\over 2} \int
d^2x \,e_2$ and  leads  to the much
studied two-dimensional $\sigma$-model whose static solutions are harmonic
maps from ${\bf R}^2$ to $S^2$. However, the $\sigma$-model is not a good
analogue of the Skyrme model  since, unlike Skyrmions,  its soliton solutions
have an arbitrary scale.
The scale invariance is broken by adding the Skyrme term ${1\over 2}\int d^2x
\,e_4$,
but the resulting energy functional is still not satisfactory: it can
have no minima since the energy of  any configuration can be lowered
by rescaling: $\bphi(\bx) \rightarrow \bphi({\bx \over \lambda})$, where
$\lambda >1$.
Hence it is necessary to include a term in the energy functional which
contains no derivatives of the field $\bphi$ and which  is often called
a potential. Here we will consider  the potential $\mu^2(1-\bn\cd\bphi)$, where
$\bn =(0,0,1)$ and $\mu$ is a constant with the dimension of inverse length.
Our potential is analogous to the extra term included in the Skyrme model to
give the pions a mass (thus $1/\mu $ may physically be interpreted as the
Compton wavelength of  the mesons in our model). Both in the Skyrme model  and
in our model the potential
term  reduces the symmetry of the model and is responsible for the soliton
solutions being exponentially localised
in space.
In  the Skyrme model  the size of Skyrmion is slightly smaller than the pion's
Compton wavelength, reflecting  the relative magnitudes of  nucleon size and
the pion's Compton wavelength in nature. To mimic these properties we
want to have a basic soliton  solution whose size is of order $1/\mu$. This can
be achieved by setting $\mu^2 =0.1$, which we do for the rest of this paper.
Note, however, that a  different choice of $\mu$
results in a different model. While some of the general features to be
discussed
in this paper are independent of $\mu$,  other
 properties, such as the shape  of multisolitons (we give a precise definition
furhter below), may well depend on it.

 Some readers may  want ot bear in mind  an alternative physical
interpretation  the field $\bphi$: it can also be thought of
as the magnetisation vector of a two-dimensional ferromagnetic substance
\cite{BP}.
Then the potential term describes the  coupling of the magnetisation vector
to a constant external magnetic field.

To sum up:  the general framework for   Skyrme models described above,
augmented by a potential modelled on the one used in the usual Skyrme model
leads to the following energy density for  the baby Skyrme model:
\bea
\label{end}
 e={1\over 2}(\partial_1\bphi^2 +\partial_2\bphi^2)
+{1\over 2}(\partial_1\bphi \times \partial_2\bphi)^2 + \mu^2
(1-\bn\cd\bphi),
\eea
from which the energy functional is obtained by integration
\bea
\label{energy}
E [\bphi]= \int d^2x\, e.
\eea
We  are only  interested in configurations with finite energy, so we define our
configurations space $Q$ to be the space of all maps $\bphi:{\bf R}^2\mapsto
S^2$
which tend to the constant field $\bn$, called the vacuum, at spatial infinity
\bea
\label{bound}
\lim_{|\bx| \rightarrow \infty} \bphi(\bx) = \bn.
\eea
As a result,  a configuration $\bphi$ may be regarded as a map from
compactified
physical space ${\bf R}^2\cup \{ \infty \}$, which is homeomorphic
 to a 2-sphere,
to $S^2$. Thus every configuration $\bphi$ may be regarded as a representative
of a homotopy class in $\pi_2(S^2)={\bf Z}$ and has a  corresponding integer
degree,
which can be calculated from
\bea
\mbox{deg}[\bphi] = {1\over 4 \pi}\int   d^2x\,\bphi \cd \partial_1 \bphi
\times
\partial_2 \bphi .
\eea
Configurations with different
degrees cannot be smoothly deformed into each other and hence  the
configuration  space $Q$ is not
connected. We write $Q_n$ for the component of $Q$ containing the
configurations of degree $n$.
 In this paper we are interested in stationary points and,
if they exist, minima of $E$ in a given
sector $Q_n$. A configuration $\bphi$ is a  stationary point
 of $E$ if  the  first variation  of $E$  under
$\bphi(\bx)\mapsto \bphi(\bx) + \bep(\bx)\times\bphi(\bx)$  vanishes  for any
function
$\bep:{\bf R}^2\mapsto {\bf R}^3$ satisfying $\bphi(\bx)\cd\bep(\bx) =0$ for
all $\bx$. This requirement leads to the Euler-Lagrange equation
\bea
\label{EL}
\partial_i\bj_i = \mu^2\bn\times\bphi,
\eea
where
\bea
\label{current}
 \bj_i = \bphi \times \partial_i\bphi + \partial_j\bphi (\partial_j\bphi
\cd\bphi\times\partial_i\bphi).
\eea

 The degree  gives a useful lower bound
on the potential energy, the Bogomol'nyi bound
\bea
E[\bphi] \geq 4\pi \cd |\mbox{deg}[\bphi]|.
\eea
This inequality holds already for the $\sigma$-model energy functional
$E_{\sigma}$ (for
a proof see \cite{Raja} ) and since $E$ is bounded below by $E_{\sigma}$ it
holds
for $E$ as well. Assuming that there are finite energy configurations of
degree $n$
it follows that  the infimum of the  restriction  $E_{|Q_n}$ of the functional
$E$ to $Q_{n}$
 exists, and  we call it  $E_n$. It is not clear, however,  whether there is
a configuration of degree $n$ whose energy is $E_n$. If  such a configuration
exists for $n=1$ we call it  a 1-soliton  or  a  baby Skyrmion, and if it
exists  for $n>1$ and if  moreover its energy  satisfies
 \bea
\label{ineq}
E_n <E_k +E_l  \qquad \mbox{for all integers}\, 1< k,l< n\quad \mbox{such that}
\quad k+l=n
\eea
 we call it a multisoliton or an $n$-soliton. If we do not need  to specify the
degree we simply write soliton
for  an $n$-soliton with  arbitrary $n \in{\bf N}$ (all solitons of negative
degree can be obtained from solitons of positive degree via the iso-reflection
$(\phi_1,\phi_2,\phi_3)\mapsto (\phi_1,-\phi_2,\phi_3)$, so  we can restrict
attention to positive $n$ without loss of generality).
We have  included the condition (\ref{ineq}) in our definition because we
want multisolitons to be stable with respect to decay into multisolitons of
smaller degree.

 The goal of this paper is  to show  the existence of multisolitons in the baby
Skyrme model  and  to describe their properties. In section 2 we discuss
a particular ansatz for finding solitons and use it to calculate the field of a
 baby Skyrmion.
 In section 3 we set up a general framework for  proving  the inequality
(\ref{ineq}) and in section 4 we complete the proof  for certain values of
$k,l$ and $n$.
Section 5  contains  numerical  evidence  for the existence of $n$-solitons
in our model for $1\leq n\leq 6$ and a description of  their properties.

\section{Hedgehog Fields  and  Baby Skyrmions}

 A powerful  method of finding stationary points of the energy functional $E$
(\ref{energy}) exploits the invariance  of $E$ and $Q_n$ under the symmetry
group
\bea
G = E_2 \times SO(2)_{\mbox{\tiny iso}} \times P.
\eea
Here $E_2$ is the euclidean
group of translations and rotations in two dimensions which acts on
fields via pull-back.
$SO(2)_{\mbox{\tiny iso}}$ is the subgroup of  the three-dimensional rotation
group  acting  on $S^2$
 which leaves  $\bn$ fixed. We call its elements  iso-rotations to distinguish
them from rotations in physical space. Elements of $SO(2)_{\mbox{\tiny iso}}$
can be parametrised by an angle $\chi \in [0,2\pi)$ and act on $\bphi$ via
\bea
(\phi_1,\phi_2,\phi_3) \mapsto (\cos\chi \,\phi_1 +\sin \chi \,\phi_2,
-\sin\chi\,\phi_1 + \cos\chi\,\phi_2,\phi_3).
\eea
Finally  $P$ is a combined reflection in both space and the target space $S^2$:
\bea
P : (x_1,x_2) \mapsto (x_1,-x_2) \quad \mbox{and} \quad
(\phi_1,\phi_2,\phi_3)\mapsto(\phi_1,-\phi_2,\phi_3)
\eea

The vacuum field $\bn$ is invariant under the symmetry group $G$ and clearly
minimises  the energy  in $Q_0$.
 However, we are interested in stationary points  of degree $\neq 0$
and  the maximal subgroups of $G$ under which
such a field can be invariant  are labelled by a non-zero integer $n$ and
consist  of spatial rotations by some angle $\alpha\in[0,2\pi)$ and
simultaneous iso-rotation by $-n\alpha$.  Fields  invariant under such a  group
are of the form
\bea
\label{hedge}
\bphi(\bx) = (\sin f(r) \cos (n\theta-\chi), \sin f(r) \sin(n\theta - \chi)
, \cos f (r)),
\eea
where $(r,\theta)$ are polar coordinates in the $\bx$-plane,
 and $f$ is  function satisfying certain boundary conditions to be specified
below. The angle $\chi$ is also arbitrary,
but fields with different $\chi$ are related by an iso-rotation and therefore
degenerate in energy. Hence we  concentrate on the fields where
$\chi =0$.
Such fields are the analogue of the hedgehog  fields
in the Skyrme model and we also call them  hedgehog fields here. They were also
studied in \cite{Tigran} for a different value of $\mu$.

The  function $f$, which we call the profile function, has to satisfy
\bea
\label{origin}
f(0)= m\pi, \quad m\in \mbox{\bf Z},
\eea
 for the field (\ref{hedge}) to be regular at
the origin and to satisfy  the boundary condition (\ref{bound})  we set
\bea
\label{infinity}
\lim_{r\rightarrow \infty} f(r) = 0.
\eea
We can assume without  loss of generality that $m$
is positive  because changing the sign of $f$ in (\ref{hedge}) is equivalent to
 an iso-rotation by $180^{\circ}$. The restriction $\tilde E$ of $E$ to fields
of the form (\ref{hedge}) is
\bea
\label{enden}
\tilde E = 2\pi \int r dr\left( {1\over 2}f'^2 + {n^2 \sin^2 f\over 2 r^2}(1 +
f'^2)
+  \mu^2(1-\cos f) \right).
\eea
It follows from the ``principle of symmetric criticality" \cite{Palais}
that a field of the form   (\ref{hedge}) is a stationary point of the energy
functional $E$ if
$f$ is a stationary point of $\tilde E$, i.e. if $f$ satisfies the Euler
Lagrange equations first written down in \cite{Tigran}:
\bea
\label{rad}
\left(r + {n^2 \sin^2 f\over r}\right)f'' +\left(1-{n^2 \sin^2f\over r^2} +
{n^2 f'\sin f \cos f \over r}\right)f' -{n^2\sin f \cos f \over r} - r\mu^2\sin
f
=0.\nonumber \\
{}.
\eea
The behaviour of solutions of this equation near the origin and for large
$r$ can be deduced analytically and was also discussed in \cite{Tigran}.
The result is that, for small $r$,
\bea
f(r)  \approx m\pi + C r^n,
\eea
where $C$ is a constant which depends on $m$ and $n$.
For large $r$, the equation (\ref{rad}) simplifies to the modified Bessel
equation
\bea
f'' + {1\over r} f' - ({ n^2\over r^2} + \mu^2)f=0.
\eea
Solutions of this equation which tend to zero at $r=\infty$ are the modified
Bessel functions $K_n(\mu r)$
of
order $n$. Thus a solution $f$ of (\ref{rad}) is proportional to $K_n$ for
large $r$ and we can write
\bea
\label{profasy}
f(r) \sim {c_n \mu^n \over2 \pi}K_n(\mu r),
\eea
where $c_n$ is a constant, dependent on $n$ and $m$, which we will
interpret further below. Since the modified Bessel functions  have the
asymptotic behaviour
\bea
\label{asympt}
K_n(\mu r) \sim \sqrt{\pi \over 2 \mu r} e^{-\mu r}\left( 1+ {\cal O}({1\over
\mu r})\right)
\eea
we know that the leading term in an asymptotic expansion of  $f$ is
$e^{-\mu r}/ \sqrt r$.
The behaviour
of $f$
at infinity guarantees that the energy distribution of the corresponding
field $\bphi$ is exponentially
localised and  hence that its total energy is finite.

What is the degree of the field (\ref{hedge})? A short calculation
gives
\[
\mbox{deg}[\bphi] =  \left\{ \begin{array}{ll}
                         n& \mbox{if $m$ odd} \\
                         0& \mbox{if $m$ even}.
                   \end{array}
\right. \]
Thus for a given degree  $n\neq 0$ there are infinitely many  solutions
of the static field equations  of the form
(\ref{hedge}), one for each odd $m$.

Using a shooting method we have solved the ordinary  differential equation
(\ref{rad}) numerically
 for a  range of values of $n$ and $m$ and computed
the energy of the corresponding hedgehog field from (\ref{enden}).
We find that, amongst all solutions of the static field equations  of the form
(\ref{hedge})
 with  degree 1 , the field
with $n=m=1$ has the lowest energy.
We write $\bphi^{(1)}$ for that field in standard orientation ($\chi=0$) and
denote the profile function
by  $f^{(1)}$.
 We have also numerically tested the stability of the field $\bphi^{(1)}$
against   perturbations which destroy the  rotational symmetry
and found  it to be stable (we will describe our numerical method in more
detail in section 5).
 Thus, in analogy to  the Skyrme model, the
 minimal  energy configuration amongst all
 fields of degree one is of the hedgehog form.
The profile function  $f^{(1)}$ is plotted in figure  1.a) and a plot of  the
energy
density $e$ of the field, which is given by the expression in round brackets in
(\ref{enden}), is shown in figure 1.b). The actual  field $\bphi^{(1)}$ is
displayed in figure 2.a).

 There is a whole  manifold ${\cal M}_1$
of minima of $E_{|Q_1}$
obtained by acting with $G$ on $\bphi^{(1)}$.
All  elements of ${\cal M}_1$  are baby Skyrmions as defined at the end  of
section 1.  ${\cal M}_1$ is three-dimensional, so   a baby Skyrmion is
characterised by its two-dimensional position vector and an angle $\chi$
specifying its orientation.
The value of $E$ on ${\cal M}_1$ physically represents the  mass $E_1$ of a
baby Skyrmion. Numerically
we find $E_1=1.564 \cd 4\pi$.

\section{Existence of  Multisoliton Solutions}

The basic ingredient for proving (\ref{ineq}) is a  superposition procedure for
solitons. In our model such a procedure  can be found using
the stereographic projection
\bea
p: {\bf C}\cup\{\infty\}&\mapsto& S^2 \nonumber \\
   u=u_1+iu_2  &\mapsto& ({2u_1 \over 1+|u|^2},{2u_2 \over 1+|u|^2},
{1-|u|^2\over1+|u|^2}).
\eea
This allows us to translate a function
\bea
u: {\bf R}^2\cup\{\infty\}\mapsto {\bf C}\cup\{\infty\} , \qquad u(\infty)=0
\eea
 into a configuration $\bphi^u$ via
\bea
\bphi^u=p\circ u
\eea
and vice versa. If we think of  $u$ as  a function of $z=x_1+ix_2$ and
$\bar z =x_1-ix_2$ then its degree, which equals the degree of $\bphi^u$,  is
the number of poles
in $z$, counted with multiplicity, minus the number of poles
in $\bar z$, also counted with multiplicity. For more details on the
formulation
 of  two-dimensional Skyrme  models in terms of $u$ instead of
$\bphi$ we refer the reader to \cite{CP1pot}.  Here we use it to  define: the
superposition of  the configurations $\bphi^u$ and
$\bphi^v$
 is  the configuration $\bphi^w$ where $w=u+v$. Then clearly
$\mbox{deg}[\bphi^w]=\mbox{deg}[\bphi^u]+ \mbox{deg}[\bphi^v]$.

Now let $\bphi^u$ and $\bphi^v$ be multisolitons of degrees $k$ and $l$.
We want to show that under
certain circumstances $E[\bphi^w]<E[\bphi^u]+E[\bphi^v]$.
The idea for our proof is taken from an unpublished paper by Kugler and
Castillejo \cite{Kugler} where the authors claim to to prove the corresponding
statement for the Skyrme model. However, their proof contains an unjustified
assumption as we shall see further below.

Consider
the situation where the multisolitons $\bphi^u$ and $\bphi^v$ are well
separated.
More precisely we assume that we can divide ${\bf R}^2$ into two regions such
that $u$ is small in region 2 and $v$ small in region 1. The smallness of
$u$ means that $\bphi^u$ is close to the vacuum  $\bn$
in region 2 and hence  of the form
\bea
\bphi^u =\sqrt{1-\bvphi^u\cd\bvphi^u}\bn +\bvphi^u\approx\bn+\bvphi^u +{\cal
O}((\bvphi^u\cd\bvphi^u),
\eea
where $\bvphi^u\cd \bn=0$.
Then, if $\bphi^u$ satisfies the Euler-Lagrange equation (\ref{EL}), $\bvphi^u$
satisfies the linearised equation
\bea
(\Delta-\mu^2)\bvphi^u =0
\eea
in  region 2.
Similarly, $\bphi^v$ is close to the vacuum in region 1 and one defines
$\bvphi^v$
analogously to $\bvphi^u$. It satisfies
\bea
\label{reg1}
(\Delta-\mu^2)\bvphi^v =0
\eea
in  region 1.
One checks that $\bphi^w$  has the following expansion in powers of $\bvphi^v$
in region 1
\bea\bphi^w\approx  \bphi^u + \bep^v\times \bphi^u +{1\over 2}
\bep^v\times(\bep^v\times \bphi^u),
\eea
where $\bep^v$ is  linear in $\bvphi^v$ but also depends  non-linearly on
$\bphi^u$:
\bea
\bep^v={1\over 2} \bphi^u\times \left( (1+\bphi^u\cd\bn)\bvphi^v -(\bphi^u\cd
  \bvphi^v)\bn) \right).
\eea
Similarly, in region 2
\bea
\bphi^w\approx  \bphi^v + \bep^u\times \bphi^v +{1\over 2}
\bep^u\times(\bep^u\times \bphi^v),
\eea
where $\bep^u$ depends linearly on $\bvphi^u$ but non-linearly on $\bphi^v$:
\bea
\bep^u={1\over 2} \bphi^v\times \left( (1+\bphi^v\cd\bn)\bvphi^u -(\bphi^v\cd
  \bvphi^u)\bn) \right).
\eea

Using these formluae we evaluate $E[\bphi^w]$: in region 1 we  keep
terms which
involve $\bphi^u$ only, terms which are linear in $\bvphi^v$ and  those terms
quadratic in $\bvphi^v$ which are independent of $\bphi^u$. Similarly, in
region 2 we keep terms which involve $\bphi^v$ only, terms  which are linear
 $\bvphi^u$  and those terms quadratic in $\bvphi^u$ which are independent of
$\bphi^v$.
Denoting integration over region 1 and 2 simply by the suffix 1 and 2
respectively we find

\vbox{
\bea
E[\bphi^w]&\approx& \int_1 d^2x \,e\left(\bphi^u + \bep^v\times \bphi^u
+{1\over 2}
\bep^v\times(\bep^v\times \bphi^u)
\right) \nonumber \\
&+&
\int_2 d^2 x\, e \left( \bphi^v + \bep^u\times \bphi^v + {1\over 2}
\bep^u\times(\bep^u\times \bphi^v) \right) \nonumber \\
&\approx &
\int_1d^2x \,e(\bphi^u)  +  \int_2 d^2 x
 \left({1\over 2}\partial_i\bvphi^u\cd \partial_i \bvphi^u
+{1\over 2}\mu^2\bvphi^u\cd\bvphi^u\right)
\nonumber\\
&+&  \int_2 d^2x\, e(\bphi^v) + \int_1 d^2 x \left({1\over
2}\partial_i\bvphi^v\cd \partial_i \bvphi^v
+{1\over 2}\mu^2\bvphi^v\cd\bvphi^v\right)
\nonumber \\
&+&  \int _1 d^2x\,\bj^u_i\cd \partial_i\bep^v + \mu^2
\bep^v\cd\bn\times\bphi^u
+  \int _2 d^2x\,\bj^v_i\cd\partial_i\bep^u + \mu^2 \bep^u\cd\bn\times\bphi^v,
\eea}

\noindent where $\bj^u_i$  and $\bj^v_i$ denote the current $\bj_i$
(\ref{current}) evaluated on the fields
$\bphi^u$ and $\bphi^v$ respectively.
The integrals in the third  line represent  the   contribution to energy  of
the  soliton $\bphi^u$  from  region 1 and  the leading contribution from
region 2. The integrals
in the third line are the  contribution to the energy of $\bphi^v$ from region
2 and the leading contribution from region 1. The formulae for subleading
contributions are complicated,
but it is clear that the sum of all terms which only involve either $\bphi^u$
or
$\bphi^v$ is just $E[\bphi^u] + E[\bphi^v]$. The
cross terms are more interesting. Integrating by parts in the last line of the
equation above, and using the fact that both $\bphi^u$ and $\bphi^v$ satisfy
the Euler-Lagrange equations (\ref{EL}), we are only left with  a boundary term
\bea
E[\bphi^w] \approx E[\bphi^u]+E[\bphi^v] + \int_{\Gamma} (\bj_i^v\cd \bep^u -
\bj_i^u\cd\bep^v)dS_i.
\eea
Here $\Gamma $ is a curve without self-intersections  separating the region 1
from the region 2 and
$dS_i=\varepsilon_{ij}\dot \gamma_j dt$ for any parametrisation of $\gamma(t)$
of $\Gamma$ for which  region 1 is on the left and region 2 on the right
($\varepsilon_{ij}$ is the antisymmetric tensor in two dimensions normalised so
that $\varepsilon_{12} =1$).
We now assume further that $\Gamma$ lies in a region where both $\bphi^u$
and  $\bphi^v$ are close to the vacuum and keep only terms which are linear
in both $\bvphi^u$ and $\bvphi^v$. Then, interpreting the difference $
E[\bphi^w] -E[\bphi^u]-E[\bphi^v] $ as the potential describing the interaction
of the solitons $\bphi^u$ and $\bphi^v$ we find that the leading term $V$
of that  potential
is given by the simple formula
\bea
\label{asy}
V = \int_{\Gamma}( \bvphi^v\cd\partial_i\bvphi^u -\bvphi^u\cd\partial_i\bvphi^v
)dS_i.
\eea

To prove (\ref{ineq}) one needs to show that we can always arrange for $V$ to
be negative. Suppose that this is not already the case, and consider the case
where $V$ is positive. Then it can be made negative by an iso-rotation by
$180^{\circ}$
of either $\bphi^u$ or $\bphi^v$ (which does not change the individual energies
of  $\bphi^u$  and
$\bphi^v$). However, the proof is incomplete unless we can rule out that
$V$ is zero. This possibility was not considered in \cite{Kugler}. In fact  we
shall see that this does indeed
happen for all  relative iso-orientations in a slightly modified version of our
model. Thus we can only use  the result
(\ref{asy}) for constructing new multisolitons out of solitons  for
which the asymptotic field $\bvphi$ is known.

\section{New  Multisolitons from Old}

Although we  already have enough information about the asymptotic field of a
baby Skyrmion to prove
the existence of a 2-soliton in our model, it is useful  to interpret the
asymptotic field further before proceeding with the proof.

For large $r$ the profile function $f^{(1)}$
 approaches $0$ exponentially fast and we can therefore approximate
$\sin f^{(1)} \sim f^{(1)}$ and $\cos f^{(1)} \sim 1$.
Using the asymptotic expression (\ref{profasy}) we write the
asymptotic form $\bvphi^{(1)}$ of $\bphi^{(1)}$ as
\bea
\bvphi^{(1)}(\bx) = {p\mu\over 2\pi}(K_1(\mu r)\cos (\theta-\chi) ,
  K_1(\mu r)\sin(\theta -\chi),0),
\eea
where we have written $p$ for $c_1$.
Alternatively, introducing the orthogonal  vectors
\bea
\bp_1 = p(\cos\chi,\sin\chi) \qquad \bp_2= p(-\sin\chi, \cos \chi)
\eea
and $\hat \bx = \bx/r$ we can write
\bea
\varphi^{(1)}_a(\bx)={\mu \over  2\pi} \,\bp_a\cd\hat\bx K_1(\mu r)
= -{1\over 2 \pi}
\bp_a\cd\nabla K_0(\mu r)
 \qquad a=1,2.
\eea
Now, given  that the Green function  of the static
Klein-Gordon equation in two dimensions is
$K_0(\mu r)$:
\bea
\label{Green}
(\Delta -\mu^2)K_0(\mu r)= -2\pi \delta^{(2)}(\bx),
\eea
it follows that
\bea
(\Delta -\mu^2)\varphi_a^{(1)}(\bx) = \bp_a\cd\nabla\delta^{(2)}(\bx)
\qquad a=1,2.
\eea
Thus the asymptotic field $\bvphi^{(1)}$ may be thought of  as
produced by a  pair
of orthogonal dipoles, one for each of the components $\varphi^{(1)}_1$ and
$\varphi^{(1)}_2$, in a linear field theory, namely Klein-Gordon theory.
The strength of the dipole can be calculated from the asymptotic form
of $f^{(1)}$. We find
\bea
p = 24.16. \qquad
\eea

Consider now the set up of section 3, with $\bphi^u$ a baby Skyrmion centred at
 the origin and iso-rotated relative to the standard hedgehog field
$\bphi^{(1)}$  by $\chi_1$  and $\bphi^v$ a second baby Skyrmion  iso-rotated
relative to the standard hedgehog by an angle $\chi_2 $ and centred at $\bR$,
where $R:=|\bR|<<1/\mu$. The asymptotic field of  the first baby Skyrmion
 is
\bea
\label{standard}
 \varphi^u_a(\bx)= -{1\over 2 \pi}\bd_a\cd\nabla K_0(\mu r)
 \qquad a=1,2.
\eea
 where
\bea
\label{dips}
\bd_1 = p(\cos \chi_1,\sin \chi_1) \qquad \bd_2= p(-\sin \chi_1, \cos \chi_1),
\eea
and the asymptotic field of the second is
\bea
 \varphi^v_a(\bx)= -{1\over 2 \pi}\bp_a\cd\nabla K_0(\mu |\bx-\bR|)
 \qquad a=1,2,
\eea
 where
\bea
\bp_1 = p(\cos\chi_2,\sin\chi_2) \qquad \bp_2= p(-\sin\chi_2, \cos \chi_2).
\eea
Then, using that, in region 1,
\bea
\label{source}
(\Delta -\mu^2)\varphi^u_a =  \bd_a\cd\nabla\delta^{(2)}(\bx)
 \quad \mbox{and}\quad (\Delta -\mu^2)\varphi^v_a =0  \qquad a=1,2
\eea
and converting the  line integral (\ref{asy}) into an area integral over
region 1
we find the potential for the  interaction of two well separated
baby Skyrmions
\bea
\label{11}
V_{11}&=& \int_1 d^2 x\, \bvphi^v \cd (\Delta -\mu^2)\bvphi^u
\nonumber \\
&=& \sum_{a=1,2}{1\over 2 \pi}
(\bd_a\cd\nabla)(\bp_a\cd\nabla)K_0(\mu R)\nonumber\\
 &=& {p^2\over \pi} \cos \psi \Delta K_0(\mu R)\nonumber \\
& =&
{p^2 \mu ^2 \over \pi}\cos \psi K_0(\mu R),
\eea
where $\psi =\chi_1 -\chi_2$  describes the relative iso-orientation.
In the last step we have used (\ref{Green}) and  have omitted the
$\delta^{(2)}$-function term   because we are only interested in large
separations $R>1/\mu$. Thus we see that the potential for the interaction of
two well
separated baby Skyrmions is the same as that calculated in a linear field
theory,
namely Klein Gordon theory for a pair of scalar fields, for the interaction
of two pairs of orthogonal dipoles. The important  point is that baby Skyrmions
not only act  as sources of dipole fields but also react to an
external field like a pair of orthogonal dipoles. This result deserves some
comments. Firstly, it is instructive to  compare it
 to a similar result in the Skyrme model. There  a certain non-linear
superposition procedure, the product ansatz, also
leads to  an interaction potential  which has a ``linear" interpretation:
it describes the interaction between two  triplets of mutually orthogonal
scalar dipoles in a linear field theory for the pion fields,
see \cite{Skyrprod} and also \cite{S}.

Secondly, we are now in a  position to explain the
caveat at the end of the previous section  concerning the use of (\ref{asy})
for proving (\ref{ineq}). In \cite{CP1pot} a similar model to ours was studied
with the potential term $\mu^2(1-\bphi\cd \bn)$ replaced by
$(1-\bphi\cd\bn)^4$.
In that model (where the mesons are massless) the solitons of degree 1 also
have the hedgehog form (\ref{hedge}) (with a particularly simple profile
function)  and  their
asymptotic field can still be interpreted in terms of a pair   of dipoles.
The Green function of the linearised theory, however, is $-{1\over 2\pi} \ln
R$. Thus a calculation analogous to ours yields a  potential proportional
to $\cos \psi \Delta \ln R$  which is identically zero for all values of
$\psi$. Numerically, two solitons  of degree one are  found to repel each other
so that there are no 2-solitons in that model.  The example shows that the
``linear forces"  between solitons, calculated via (\ref{asy}), may vanish for
all relative iso-orientations. In that situation the inter-soliton forces
are entirely due to non-linear effects and cannot be calculated with the
methods of the previous section.

Returning to the formula (\ref{11})  we see that    $V_{11}$
is negative if one baby Skyrmion  is iso-rotated by $180^{\circ}$
relative to the other, i.e. if $\psi =\pi$ in the above expression.
Thus we conclude that  $E_2 < 2E_1$. In fact one finds  that  already the
minimum of the
energy amongst hedgehog fields of degree 2 is less than $2 E_1$.
 We write   $\bphi^{(2)}$ for   the hedgehog field
(\ref{hedge})  with  $n=2$ in standard orientation ($\chi=0$) whose profile
function  satisfies  (\ref{infinity}),(\ref{rad}) and (\ref{origin})
with $m=1$. For its energy, or mass,  we find $E_2=2.936 \cd 4\pi$.
We have again checked the stability of $\bphi^{(2)}$
against more general perturbations numerically  and
conclude that it minimises the energy amongst all fields of degree $2$.
Its profile function is plotted in figure 1.a) and its energy density in figure
 1.b).
Note that the maximum of the energy density is not at the origin but at
$r\approx 1.8$. This is again reminiscent of the Skyrme
model, where the energy of the static solution of degree $2$ is
concentrated in a toroidal region \cite{axi}. The field $\bphi^{(2)}$ is shown
in figure  2.b).

Before we can use the 2-soliton as an input for the construction
of higher multisolitons we need to understand the asymptotic form
$\bvphi^{(2)}$ of the field $\bphi^{(2)}$. Since the 2-soliton
is of the hedgehog form this is not difficult. The asymptotic field
is now
\bea
\bvphi^{(2)}={q\mu^2\over 2\pi}(K_2(\mu r)\cos(2\theta -\chi),K_2(\mu r)
\sin(2\theta-\chi),0)
\eea
and can be expressed in terms of the quadrupole moments
\bea
\bq_1=q(\cos \chi,\sin\chi)\qquad \bq_2=q(-\sin \chi,\cos \chi)
\eea
and the second order differential operator
$\bD=(\partial_1^2-\partial_2^2,2\partial_1 \partial_2)$ via
\bea
\varphi^{(2)}_a ={1\over 2\pi}\bq_a\cd\bD K_0(\mu r)\qquad a=1,2.
\eea
Then, using again (\ref{Green}),
it follows that
\bea
(\Delta-\mu^2)\varphi^{(2)}_a=-\bq_a\cd\bD\delta^{(2)}(\bx).
\eea
Thus we  may think of the asymptotic field  $\bvphi^{(2)}$ as being due
to a pair of orthogonal quadrupoles (in two dimensions all
multipoles have two real components). For the strength of the quadrupole
we find
\bea
q= 53.6.
\eea

Now consider the superposition of a baby Skyrmion
at  the origin and iso-rotated relative to the standard hedgehog by $\chi_1$
and a 2-soliton centred at $\bR$, $R>>1/\mu$, and
iso-rotated relative to the standard hedgehog $\bphi^{(2)}$ by $\chi_2$.
Thus the asymptotic field and the dipole moments of the baby Skyrmion
are as in (\ref{standard}) and (\ref{dips})
and the asymptotic field of the second is
\bea
 \varphi^v_a(\bx)= {1\over 2 \pi}\bq_a\cd\bD K_0(\mu |\bx-\bR|)
 \qquad a=1,2,
\eea
 where
\bea
\label{quad}
\bq_1 = q(\cos\chi_2,\sin\chi_2) \qquad \bq_2= q(-\sin\chi_2, \cos \chi_2).
\eea
Then, inserting these expressions into the general formula (\ref{asy}) and
using (\ref{source})
we find  the potential $V_{12}$ describing the interaction between a baby
Skyrmion and a 2-soliton :
\bea
V_{12} &=& \sum_{a=1,2}{1\over 2 \pi}
(\bd_a\cd\nabla)(\bq_a\cd\bD)K_0(\mu R)\nonumber\\
& =&{pq\over 2\pi}(\cos\psi \,\partial_1 + \sin\psi\, \partial_2)\Delta K_0(\mu
R)\nonumber \\
&=&-{pq \mu ^3 \over 2\pi}\cos (\psi -\vartheta) K_1(\mu R),
\eea
where $(R,\vartheta)$ are polar coordinates for the relative position vector
$\bR$ and $\psi$ is defined as before. Thus the interaction between a baby
Skyrmion and
a 2-soliton depends both on  their relative iso-orientation and
on their relative position in physical space. By choosing $\psi-\vartheta=0$
we can again arrange for the potential to be negative, so we expect there to
be a 3-soliton solution in our model. Before describing  that solution in the
next section we calculate  the potential $V_{22}$ for the interaction between
two  2-solitons.

Thus we look at  the superposition of a 2-soliton
at the origin and iso-rotated relative to the standard hedgehog field by
$\chi_1$ and a second 2-soliton centred at $\bR$ and iso-rotated
relative to the standard hedgehog field by $\chi_2$.
For the former the quadrupole moments  are
\bea
\be_1 =q(\cos \chi_1,\sin \chi_1) \qquad \be_2 =q(-\sin \chi_1,\cos \chi_1)
\eea
and for the latter the quadrupole moments are  $\bq_1$ and $\bq_2$ as defined
in (\ref{quad}).
Thus, the interaction potential is
\bea
V_{22} &=& \sum_{a=1,2}{1\over 2 \pi}
(\be_a\cd\bD)(\bq_a\cd\bD)K_0(\mu R)\nonumber\\
& =&
-{q^2 \over \pi}\cos \psi \Delta^2 K_0(\mu R) \nonumber \\
&=& -{q^2 \mu^4 \over \pi}
\cos \psi K_0(\mu R).
\eea
Thus $V_{22}$ has  the opposite  sign from  $V_{11}$,  but the same functional
dependence on  the separation $R$ and the relative iso-orientation $\psi$.
This deserves a comment, as it may seem surprising that  the dipole-dipole
potential $V_{11}$ does not necessarily dominate
over the quadrupole-quadrupole potential $V_{22}$ at large $R$. However, it is
characteristic of linear field theories with an exponentially decaying Green
function that the leading term in  a field produced by  an $n$-pole is
independent of $n$. This observation will be important for
us later, when we study multisolitons  whose fields are not of the simple
hedgehog
form; it means that
all multipoles in the expansion of the asymptotic field are potentially equally
important when studying the interaction of such multisolitons.

For now the most important feature of $V_{22}$ is that it  is negative when
$\psi =0$. Thus we conclude  $E_4< E_2 + E_2$. However, without
further insights
into the properties of the 3-soliton we cannot  say anything about the
relative size of $E_4$ and $E_1 + E_3$.

\section{Numerical Results for Higher Multisolitons}

There are a number of ways  to   search numerically  for stationary points
of energy functionals and to check  whether a given stationary point is a
local minimum. It is very hard, however, to  ascertain whether local minima
found in this
way are global minima. We are similarly not  able to prove that
the configurations to be  described in this section are multisolitons in
the strict sense of the definition at the end of section 1. Instead
we will consider a numerically found  configuration to  be a multisoliton if it
satisfies
certain  numerical checks for a local minimum  and the
inequality (\ref{ineq}).

Here we look for stationary points of the energy functional $E$ by solving a
suitable   time-dependent
 equation  which reduces to (\ref{EL}) in the static
limit.  The time evolution according to that  equation should stop
 at stationary points of $E$.   The  equation we  use is the Lorentz covariant
equation
\bea
\label{num}
 \partial^{\alpha}(\bphi \times \partial_{\alpha}\bphi -\partial^{\beta}\bphi
(\partial_{\beta}\bphi\cd\bphi\times\partial_{\alpha}\bphi)=
\mu^2\bphi\times\bn ,
\eea
where $\bphi$ is now a function of $x^{\alpha}=(t,\bx)$ and  the indices
$\alpha,\beta=0,1,2$ are raised and lowered with the
Lorentzian metric
diag$(1,-1,-1)$,
with  an added  friction term  to  absorb
the kinetic energy.
To solve the resulting equation numerically we use a finite difference
scheme  to evaluate the space derivatives and integrate the time evolution
using the 4th order Runge-Kutta method. We use  a square grid of $200 \times
200$
points and set the time increment $dt$ to half the length of a lattice site.
 Most of our simulations are  performed
on a grid extending in both the $x^1$ and $x^2$ direction from $-20$ to $20$.

As the initial configuration we take a particular stationary point of $E$,
namely the hedgehog field
(\ref{hedge})  of degree $n$ with the profile function satisfying
the boundary condition (\ref{origin}) for $m=1$  and
 the  ordinary differential equation (\ref{rad}).
We then  add a  small perturbation
which breaks the symmetry of the hedgehog.
We have already noted that the hedgehog fields for $n=1$ and $n=2$ are
stable against such perturbations. We have  also investigated   the cases
$3\leq n \leq 6$: in those cases the
hedgehog fields are unstable  and the time evolution    ends at less symmetric
configurations with lower energy.
 These final configurations are  stable with respect to  further perturbations,
 and  their energies satisfy the inequality (\ref{ineq}) so we take   them to
be  multisolitons.

 From the  energies $E_n$ of those multisolitons  we calculate
the ``ionisation energies" $\Delta_{kl}$ defined via
\bea
\Delta_{kl} :=  E_k+E_l - E_n , \mbox{where} \quad 1\leq k,l\leq n, \quad
k+l=n.
\eea
In table 1 we summarise our
results.
All the energy values listed in the table are obtained by integrating the
solitons' energy density over the grid. The resulting  values for  $E_1$ and
$E_2$  are   $1 \% $ smaller than   the more accurate values given  earlier in
the paper, which were obtained by solving the ordinary differential equation
(\ref{rad}) and integrating (\ref{enden}). For the calculation of the
ionisation energies,
however, it is important that all  energies are calculated  in the same way.

 \vspace{0.7cm}

\centerline{
\begin{tabular}{|c|c|c|c|}
\hline
&&&\\
   $\qquad n \qquad $  &  $\quad E_n/4\pi\quad $  &  $\quad k+l \quad $  &
$\quad \Delta_{kl}/4 \pi\quad$ \\
&&&\\
\hline
&&&\\
$1$   &  $1.549 $       & -       & - \\
&&&\\
\hline
&&&\\
$2$  &  $2.907$ &   $1+1  $     & 0.191    \\
&&&\\
\hline
&&&\\
$3$ & $4.379$   &   $2+1$  & $0.077 $   \\
&&&\\
\hline
&&&\\
$4$ & $5.800$   &   $3+1$  & $0.128 $   \\
 &    &   $2+2$  & $0.014 $   \\
&&&\\
\hline
&&&\\
$5$ & $7.282$   &   $4+1$  & $0.068 $   \\
  &   &   $3+2 $  & $0.005 $   \\
&&&\\
\hline
&&&\\
$6$ & $8.693$   &   $5+1$  & $0.138 $   \\
  &   &   $4+2 $  & $0.015 $   \\
  &   &   $3+3 $  & $0.066 $   \\
&&&\\
\hline
\end{tabular}
}
\vspace{0.7cm}
\nopagebreak
\centerline{{\bf Table }1}
\centerline {\sl Multisoliton energies and ionisation energies}
\vspace{0.7cm}

 The results for $n\geq 3$
deserve a more detailed discussion.
For $n=3$  our numerical procedure leads to the configuration displayed in
figure 3.a).
A plot of the energy
density $e$ (\ref{end}) as a function of position is shown  in figure 3.b).
Unlike the solutions discussed so far the energy of the 3-soliton
is not rotationally symmetric. Instead the configuration is like a linear
molecule made up of three (distorted) baby Skyrmions aligned so that any two
neighbours
are in the most attractive relative orientation.

In principle it is still possible to analyse the asymptotic field of the
3-soliton in terms of multipole moments. In practice the absence of
a continuous symmetry makes the analysis hard and we know from the previous
section that we may have to consider many multipole moments.
We have  therefore made no attempt at  deriving the  potential for the
interaction
between 3-solitons and other multisolitons in our model and rely on
numerical evidence for  showing the existence of  higher multisolitons.

The field  and the energy density  for the 4-soliton are plotted in
figure 4.
The 4-soliton is again like a linear molecule, but this time
made up of two 2-solitons. The plot of the field shows  that the two
2-solitons have the same iso-orientation, as  expected from our
discussion of the potential $V_{22}$. The picture suggests, and table 1
confirms, that it costs very little  energy to break the 4-soliton into
2-solitons.

The field for the 5-soliton is the least symmetric of all the multisolitons
we have studied. The field and its energy density are shown in figure 5. The
5-soliton consists of an almost undistorted 2-soliton
and 3-soliton  close together. Table 1 shows that the binding between
those constituents is  very weak.  To check whether the binding is
in fact a boundary effect we have therefore repeated  the simulation
 on a grid of
$250 \times  250$ points extending in $x^1$ and $x^2$ from $-25$ to $25$. The
result
is identical to that of the first simulation.

Finally the 6-soliton is made up of  three 2-solitons
centred at the vertices of an equilateral triangle. The plot of the field
in figure 6.a) reveals  that the 2-solitons all have the same iso-orientation,
so that the interaction energy between any two 2-solitons is minimised.
Although the binding is again quite weak,  the distortion of the individual
2-solitons as a result of their interaction is clearly visible in the plot of
the energy density in figure 6.b).

The pictures of the energy density for $n\geq 4$ suggests  that the 2-soliton
(and to a lesser extent the 3-soliton) serves as  a  basic building block for
higher multisolitons. Table 1 provides further evidence for this
observation: the largest ionisation energy in table 1 is  $\Delta_{11}$,
showing that the 2-soliton is most strongly  bound, and
the ionisation energy $\Delta_{kl}$ for $n=k+l>2$  is least if   $k=2$ or
$l=2$,
showing  that   it is easiest to break up multisolitons in a way that produces
at
least one 2-soliton.

\section{Conclusions}

We have studied multisoliton solutions in a non-linear field theory
which  may be interpreted physically  as a two-dimensional
version  of the Skyrme model for nuclear physics.
Our main analytical result is the expression (\ref{asy}) for  the leading term
in the potential describing
the interaction  of two well separated  solitons  in terms of the asymptotic
fields of  the solitons. In those cases where the asymptotic
field of the solitons is known   the  leading term could be written
down explicitly in terms of the  multipoles associated with the solitons and
the Green function of the linearised theory.
The resulting formula can then be used to
prove the existence of  multisolitons of higher degree.

Using numerical methods we could  explicitly display  multisolitons of degree
$1\leq n\leq 6$.
It turns out that in our model solitons
of degree 1 and 2 are invariant under a $SO(2)$  action,  but that
higher multisolitons are only invariant under the action of finite groups.
This is   rather reminiscent of the Skyrme  model where solitons of
degree 1 and 2 have continuous symmetries \cite{axi}, but higher multisolitons
have only
discrete symmetries, see \cite{BT} and also \cite{LM}.

The relationship between solitons and the associated multipoles is
intriguing and deserves further study. It is quite possible that one can
prove general statements about the multipole expansion of the asymptotic
field of an n-soliton  for arbitrary $n$ which would allow one to complete the
proof of the inequality (\ref{ineq}) for general $k,l$ and $n$. This would be
of  wider interest since
 the method described in section 3  applies to any field theory with soliton
solutions
provided there is  some sort  of superposition procedure for well-separated
solitons and  the linearisation
of the theory is of a suitable form.

Finally  it would  be interesting to see to what extent the multipole
description
can  be used to understand the dynamical properties of the solitons
in our model. The form of the potential $V_{11}$ shows  that the forces
between two well separated  baby Skyrmions, like those between Skyrmions,
depend both on
 the relative separation and the relative orientation. This suggests that
the interactive dynamics of  two baby Skyrmions is  more complicated
than that of other topological solitons in two dimensions, such as ${\bf CP}^1$
lumps
\cite{lumps} or the solitons studied in \cite{CP1pot}, but also possibly more
relevant to Skyrmion dynamics in three dimensions. We are presently
investigating this point and will report on it elsewhere \cite{PSZ}.

\vspace{1cm}
\noindent {\bf Acknowledgements}

\noindent WJZ   thanks the  Nuffield Foundation for support of his visit
 to the
Centre de Recherches Math\'ematiques, Universit\'e de Montreal, Montreal,
Canada
and Pawel Winternitz for his invitation to the Centre and his
 support and  hospitality
at the Centre. BJS  thanks the Scientific and Engineering Research Council
for a Research Assistantship.

\pagebreak
\parindent 0pt
\centerline{\bf Figure Captions}
\vspace{1cm}
\centerline{\bf Figure 1}

a) Profile functions for the baby Skyrmion ($n=1$) and the 2-soliton
($n=2$).

b) Energy densities $e$ (\ref{end}) as a function of $r$ for the baby Skyrmion
($n=1$) and the 2-soliton ($n=2$).

\vspace{1cm}

\centerline{\bf Figure 2}

a) Plot of the field $\bphi^{(1)}$. At every lattice site in physical space we
plot an arrow
whose direction and magnitude is  that of $(\phi^{(1)}_1,\phi^{(1)}_2)$
(we identify the axis in the target space $S^2$ with those in physical space).
At the base of the arrow we put a `$+$' if $\phi^{(1)}_3$ is positive and a
`$\times$' if $\phi^{(1)}_3$ is negative. Thus the vacuum  is  represented
simply by a `$+$'.  The labels $x$ and $y$   refer to the first and second
component of the
vector $\bx$.

b)  Plot of the field $\bphi^{(2)}$ using the same conventions as in a).

\vspace{1cm}

\centerline{\bf Figure 3}

a) Field of the 3-soliton; conventions as for  figure 2.a).

b) Energy density of the 3-soliton in the  range $-10\leq x ,y\leq 10$.

\vspace{1cm}

\centerline{\bf Figure 4}

a) Field of the 4-soliton; conventions as for  figure 2.a).

b) Energy density of the 4-soliton  in the  range $-10\leq x ,y\leq 10$.

\vspace{1cm}

\centerline{\bf Figure 5}

a) Field of the 5-soliton; conventions as for  figure 2.a).

b) Energy density of the 5-soliton  in the  range $-10\leq x ,y\leq 10$.

\vspace{1cm}

\centerline{\bf Figure 6}

a) Field of the 6-soliton; conventions as for  figure 2.a).

b) Energy density of the 6-soliton  in the  range $-10\leq x ,y\leq 10$.

\end{document}